\documentclass[journal=mamobx,manuscript=article]{achemso}
\usepackage{amsmath, amsthm, amssymb}
\usepackage{epsfig}
\usepackage{color}

\title{Controlling the Interactions between Soft Colloids via Surface Adsorption}

\author{Sergei~A.~Egorov}
\affiliation{Department of Chemistry, University of
Virginia, Charlottesville, Virginia 22901, USA}
\author{Jaros{\l}aw~Paturej}
\affiliation{Department of Chemistry, University of North Carolina, Chapel Hill, North
Carolina 27599-3290, USA}
\alsoaffiliation{Institute of Physics, University of Szczecin, Wielkopolska 15, 70451
Szczecin, Poland}
\email{paturej@live.unc.edu}
\author{Christos~N.~Likos}
\affiliation{Faculty of Physics, University of Vienna, 
Boltzmanngasse 5, A-1090 Vienna, Austria}
\author{Andrey~Milchev}
\affiliation{Max Planck Institut for Polymer Research, D-55128 Mainz, Germany} 
\alsoaffiliation{Institut f{\"u}r Physik, Johannes Gutenberg 
Universit{\"a}t Mainz, Staudinger Weg 7, D-55099 Mainz, Germany}

\begin{document}

%\pacs {68.43.De, 83.25.-x, 82.70.Dd, 82.70.-y}

\maketitle

\begin{abstract}
By employing monomer-resolved computer simulations and analytical considerations based
on polymer scaling theory, we analyze the conformations and interactions of multiarm star polymers
strongly adsorbed on a smooth, two-dimensional plane. We find a stronger stretching of the arms
as well as a stronger repulsive, effective interaction than in the three dimensional case. In particular,
the star size scales with the number of arms $f$ as $\sim f^{1/4}$ and the effective interaction
as $\sim f^{2}$, as opposed to $\sim f^{1/5}$ and $\sim f^{3/2}$, respectively, in three dimensions.
Our results demonstrate the dramatic effect that geometric confinement can have on the effective 
interactions and the subsequent correlations of soft colloids in general, for which the conformation can be altered
as a result of geometrical constraints imposed on them.
\end{abstract}

\newcommand{\nc}{\newcommand}
\nc{\cntr}[1]{\begin{center}{\bf #1}\end{center}}
\nc{\ul}{\underline}
\nc{\fn}{\footnote}
\nc{\pref}{\protect\ref}
\setlength{\parindent}{.25in}
\nc{\stab}[1]{\begin{tabular}[c]{c}#1\end{tabular}}
\nc{\doeps}[3]{\stab{\setlength{\epsfxsize}{#1}\setlength{\epsfysize}{#2}
\epsffile{#3}}}
\nc{\sq}{^{2}}
\nc{\eps}{\epsilon}
\nc{\sig}{\sigma}
\nc{\lmb}{\lambda}
\nc{\lbr}{\left[}
\nc{\rbr}{\right]}
\nc{\lpar}{\left(}
\nc{\rpar}{\right)}
\nc{\ra}{\rightarrow}
\nc{\Ra}{\Rightarrow}
\nc{\lra}{\longrightarrow}
\nc{\LRa}{\Longrightarrow}
\nc{\la}{\leftarrow}
\nc{\La}{\Leftarrow}
\nc{\lla}{\longleftarrow}
\nc{\LLa}{\Longleftarrow}
\nc{\EE}[1]{\times 10^{#1}}
\nc{\simleq}{\stackrel{\displaystyle <}{\sim}}
\newcommand{\be}{\begin{equation}}
\newcommand{\ee}{\end{equation}}
\newcommand{\bea}{\begin{eqnarray}}
\newcommand{\eea}{\end{eqnarray}}
\newcommand{\R}{\vec{R}}
\newcommand{\bc}{\begin{center}}
\newcommand{\ec}{\end{center}}
\clearpage
%\pacs {68.43.De, 83.25.-x, 82.70.Dd, 82.70.-y}

\section{Introduction}\label{sec_intro}

Physical confinement of complex fluids in small pores, films, 
or tubes is known to induce 
significant changes in their structural and dynamical properties and
phase behavior, including the emergence of surface-related phase transitions \cite{evans90,sven.sm.2009}
the modification of bulk critical phenomena \cite{vink.pre.2006,wilms.prl.2010}, 
novel types of ordering \cite{schmidt.prl.1996,osterman.prl.2007,fornleitner.sm.2008,mario.sm.2009,buzza.prl.2011,buzza.sm.2011,nygard12}
as well as a shifting of dynamical arrest lines \cite{lang10}. 
As such, the detailed microscopic understanding of this confinement-based phenomenology
is of great importance both 
scientifically and technologically. 
Already when hard, nondeformable colloids are restricted within a slit \cite{schmidt.prl.1996,osterman.prl.2007}
or on the interface between two immiscible fluids \cite{fornleitner.sm.2008,buzza.prl.2011}, important new
mechanisms come forward which affect the effective interactions between 
the constituent particles. Common examples are the softening of the interactions
emerging from averaging-out the transversal degrees of freedom \cite{osterman.prl.2007,franosch.arXiv.2012} or the
development of strong electric dipoles due to dielectric mismatch between the
immiscible fluids forming the interface \cite{buzza.prl.2011,buzza.sm.2011}.
To date, substantial progress has
been made, both experimentally and theoretically, towards the
understanding of the behavior of hard-sphere-like fluids \cite{nygard12} as well as 
of dipolar mixtures under
confinement \cite{fornleitner.sm.2008,buzza.prl.2011}.  

Confined systems comprised of soft
particles have received substantially less
attention \cite{mario.sm.2009,prestipino11}.  
Even in the cases in which soft potentials have been considered,
a simplifying assumption has been made, namely that the confined entities
are point particles whose interaction potential does not depend on geometric
confinement. However, for a wide class of common soft particles,
ranging from polymer brushes to dendrimers and from microemulsions
to microgels, the confining walls {\it physically} deform the particle,
resulting into conformational changes that affect their interaction in substantial ways.
No attempt has been undertaken to-date to take this important aspect
into account and to then proceed towards an investigation of the
structure of soft, deformable particles in the
borderline between two and three dimensions. 
At the same
time, soft systems possess important unique properties due to
their inherent tunability. Thus, the effective interaction potential
between two star polymers is known to change significantly in going
from bulk (three dimensions) to  planar confinement 
(two dimensions) \cite{benhamou03}. Therefore, this tunability presents
both advantages and additional challenges. 

The central goal of the present work is to study 
structural properties of star
polymers under confinement, where the 
confined conditions are created through adsorption of the
star polymers on a planar wall, with the degree of confinement 
controlled via the star-wall adsorption strength.
The behavior of star polymers in the vicinity of a substrate is
important for a variety of scientific and technological
applications. For example, formation of colloidal aggregates with
desired ordered structure on a substrate can be achieved by
controlling effective colloidal interactions via the use of star
polymers which act as depletants. As such, adsorption of star polymers
has been actively studied both experimentally \cite{glynos07,glynos11} and 
theoretically \cite{ohno91a,ohno91b,halperin91,joanny96,kon.sm.2007,kritikos08,chremos10}.

The present study is motivated by the recent work that has
demonstrated a dramatic amplification of tension along the backbone of
branched macromolecules as a result of their adsorption on a 
substrate \cite{panyukov09,panyukov09b,milchev11,paturej12}.
By analogy, one would expect
that adsorption on a surface would result in an enhancement of
repulsive interaction between the two star polymers \cite{benhamou03}. 
In the work at hand,
this conjecture is confirmed and quantified 
both via Molecular Dynamics (MD)
simulations and scaling theoretical calculations. In particular, 
based on scaling arguments, we propose a simple and accurate
functional form for describing mean force between two adsorbed stars,
which is found to be in excellent agreement with the simulation
results for all the combinations of parameter values studied. 

\section{Computational model}

We consider a three-dimensional coarse-grained model \cite{grest87} 
of  a polymer
star which consists of $f$ linear arms with one end free and the other one
tethered to a microscopic core (seed monomer) of size $R_c$,
which is of the order of the monomer size $\sigma$.   
Each arm is composed of $N$
particles of equal size and mass, connected by bonds.
Thus, the total number of
monomers in the star (excluding the core) is $fN$. 
Previous studies on star polymers \cite{sebastian.mm.2009} has shown that this model 
capture all the characteristics of the stars' conformations and interactions.
The bonded interaction between subsequent beads 
is described by frequently used Kremer-Grest model
\cite{grest86}
 $V^{\mbox{\tiny KG}}(r)=V^{\mbox{\tiny
FENE}}(r)+V^{\mbox{\tiny WCA}}(r)$, with the so-called 'finitely-extensible
nonlinear elastic' (FENE) potential: 
\begin{equation}
 V^{\mbox{\tiny FENE}}(r)=- 0.5kr_0^2\ln{[1-(r/{r_0})^2]}.
\label{fene}
\end{equation}
The non-bonded interactions between monomers are taken into account by means of the
Weeks-Chandler-Anderson (WCA) interaction, i.e., the shifted and truncated repulsive
branch of the Lennard-Jones potential, given by:
\begin{equation}
 V^{\mbox{\tiny WCA}}(r) = 4\epsilon\left[
(\sigma/ r)^{12} - (\sigma /r)^6 + 1/4
\right]\theta(2^{1/6}\sigma-r).
\label{wca}
\end{equation}
In Eqs.~(\ref{fene}) and ({\ref{wca}),
$r$ denotes the distance between the centers of two monomers (beads),
while the energy scale $\epsilon$ and the length scale $\sigma$ are chosen
as the units of energy and length, respectively. Accordingly, the remaining
parameters are fixed at the values
$k=30\epsilon/\sigma^2$ and $r_0=1.5$. In Eq.~(\ref{wca}) we have introduced the Heaviside step function
$\theta(x)=0$ or 1 for $x<0$ or $x\geq 0$. In consequence, 
the steric interactions in our model
correspond to good solvent conditions. The substrate in the present
investigation is considered simply as a
structureless adsorbing plane, with a Lennard-Jones potential acting with
strength $\epsilon_s$ in the perpendicular $z$--direction, $V^{\mbox{\tiny
LJ}}(z)=4\epsilon_s[(\sigma/z)^{12} - (\sigma/z)^6]$. 
Since in the the majority of our simulations we
considered the case of {\it strong} adsorption,
$\epsilon_s/(k_BT)=5$ with $k_B=1$ and $T=0.12\epsilon/k_{\rm B}$, the configurations of the simulated stars were 
two-dimensional. 
The chosen value $\varepsilon_s$ resulting into an adsorption energy of $5\,k_{\rm B}T$ corresponds to values in the typical
experimental range \cite{horinek}. 
For comparison and completeness,
we have also examined the case of stars confined between two parallel, repulsive walls (slit). In this case, stars
form quasi two-dimensional configurations, since small lateral excursions of the order of the slit width $H$
are allowed.
The wall-monomer was modeled by WCA-potentials given by Eq.~\ref{wca} acting only in $z$-direction and the slit width was set to
$H = 4\sigma$.
\begin{figure}[ht]
\begin{center}
\includegraphics[scale=0.26]{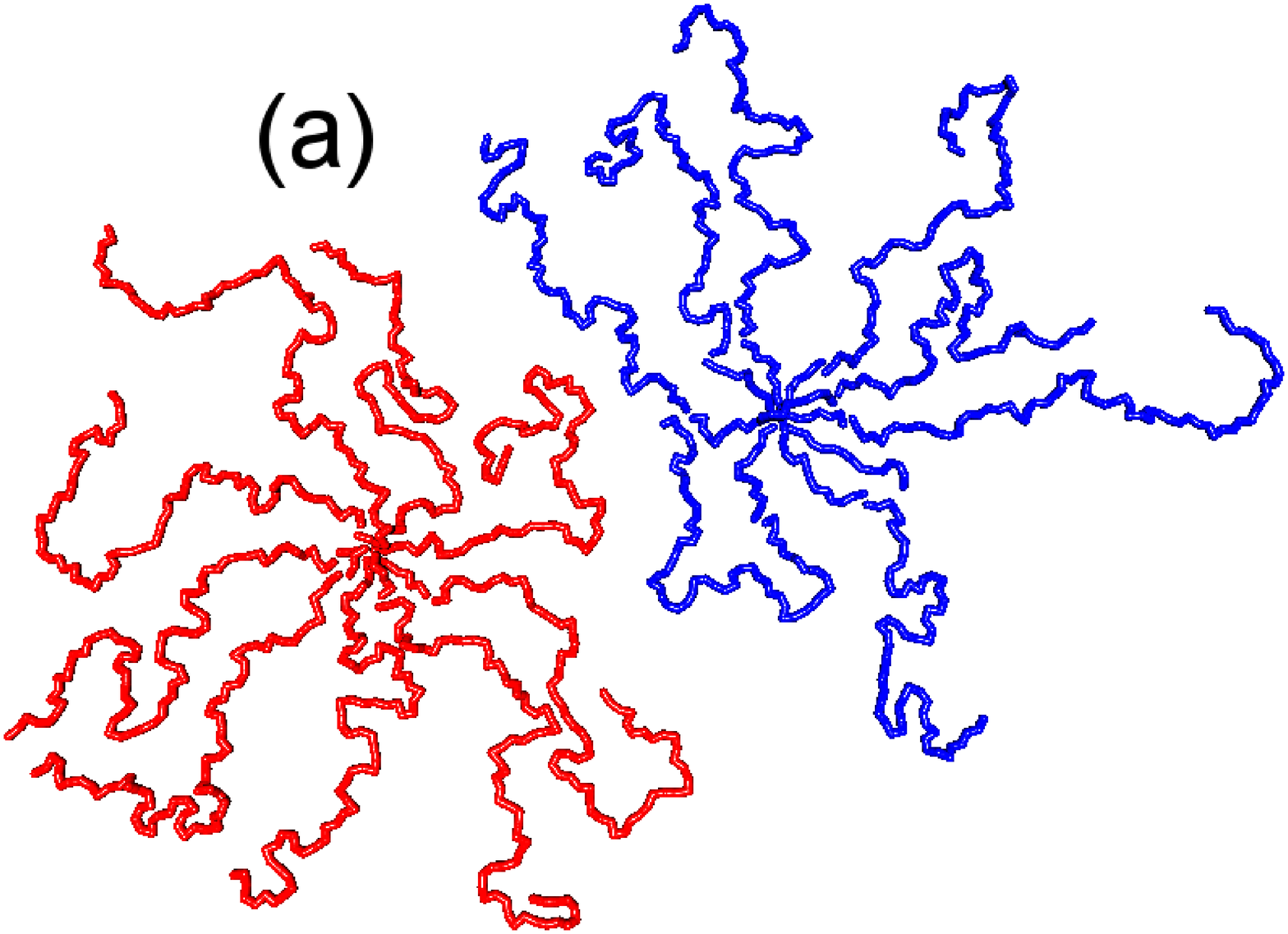}
\includegraphics[scale=0.26]{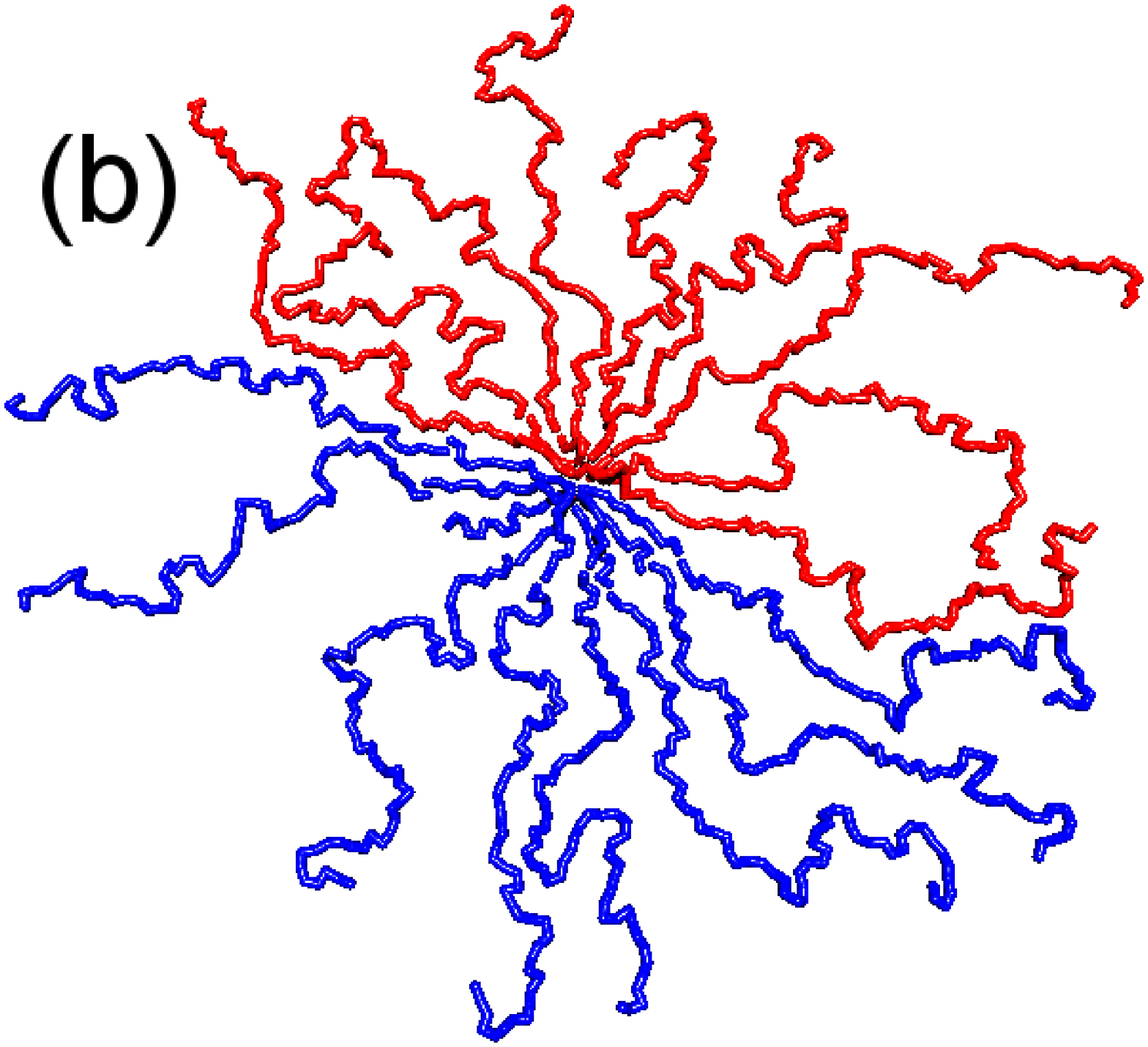}
 \caption{Snapshots of thermalized star-polymers with $f=9$ and $N=75$ adsorbed
on a solid substrate
displayed for different separation distances: (a) $d=35\sigma$; (b) $d=3\sigma$. 
}
\label{snap:fig}
\end{center} 
\end{figure}

The equilibrium dynamics of the chains is obtained 
by solving the Langevin equation of motion for
the position $\mathbf r_n=[x_n,y_n,z_n]$ of each bead in the star,
\begin{equation}
m\ddot{\mathbf r}_n = \mathbf F_n^{\mbox{\tiny FENE}} + \mathbf F_n^{\mbox{\tiny
WCA}} -\zeta\dot{\mathbf r}_n + \mathbf R_n(t), \,\,
n=1,\ldots,fN
\label{langevin}
\end{equation}
which describes the Brownian motion of a set of interacting monomers.
Even though Molecular Dynamics would be equally well-suited for gathering equilibrium statistics,
the approach based on the Langevin equation allows for the introduction of a fixed
temperature via the coupling to a random force (see below) in a natural way.
In Eq.~(\ref{langevin}) above, $\mathbf F_n^{\mbox{\tiny FENE}}$ and $\mathbf F_n^{\mbox{\tiny WCA}}$ 
are the deterministic forces exerted on monomer $n$ by the remaining bonded and nonbonded
monomers, respectively. 
The influence
of the solvent is split into a slowly evolving viscous force $-\zeta\dot{\mathbf r}_n$
and a rapidly fluctuating stochastic force. This random, Gaussian force $\mathbf
R_n$ is related to friction coefficient $\zeta$ by the fluctuation-dissipation
theorem. The integration step employed was $\Delta t = 0.002$ time units, whereby time is measured
in units of $\sqrt{m\sigma^2/\epsilon}$, $m = 1$ denoting the mass of the
beads. The ratio of the inertial forces over the friction forces in
Eq.~(\ref{langevin}) is characterized  by the Reynolds number ${\rm Re} =
\sqrt{m \epsilon}/\zeta \sigma$ which in our setup is
${\rm Re}=4$. In  the course of simulation the velocity-Verlet algorithm
was employed to integrate the equations of motion (\ref{langevin}).

Starting configurations were generated as radially straight
chains fixed to the immobile seed monomers. Stars were placed at a distance
$z=\sigma$ above the surface and at large separation distance between seeds $d$,
in order to avoid monomer overlaps between neighboring molecules. 
As a next step star-polymers were
equilibrated  until they adopted their equilibrium configurations, from which sizes
and monomer profiles have been determined.
The required equilibration time depend on $f$
and $N$, i.e., $5\times 10^5$ time steps 
for the smallest stars and up to $2\times 10^6$ for the largest.
As a next step star-polymers were equilibrated until they adopted their
equilibrium configurations. This step was controlled by monitoring 
the time evolution of the center-to-end radius $R_e$ and the
gyration radius $R_g$, as well as of the
and radial density profiles of the stars. 

Well-equilibrated configurations from a large interstar separation
were used as starting points for the next
simulation, for which we move the second seed by distance $\Delta d$ closer
to the first one and repeated the equilibration step. 
In order to sample the interaction force for each equilibrated
configuration (at a given distance $d$) a production run was performed during
which time averaged force $\langle F(t,d) \rangle = F(d)$ exerted on the
seed was computed. 
 To assure that our measured forces were true equilibrium results, we have also
employed the reverse path of first bringing the two seeds very close to one another and then
increasing their mutual distance. The resulting force vs.-distance curves turned out to be 
independent of the procedure of moving the stars close or away from each other, excluding thus
any hysteresis effects in our simulations.

During the force measurement all particles (including the cores of the stars) were allowed to fluctuate in the
$z$-direction. %but the projection of the seeds' coordinates on the adsorbing plane was kept fixed.
The fluctuations of seeded monomers in the $z$-direction do not contribute 
to the mean force $F(d)$, since the time average 
of the $z$-component of the force,
$\langle F_z(t,d) \rangle$, for seeded monomers vanishes.
The run was continued until the uncertainty of the average forces for all three
components dropped below $0.005 \epsilon / \sigma$.
Our checks have also shown that  time averaged net force acting on the seed is
equivalent to the time averaged force felt by the core of a star polymer.
Typical simulation snapshots of two interacting stars are presented in \ref{snap:fig}.

\section{Theory and comparison with simulations}

The theoretical approach we adopt for describing both single-star conformational properties
and the effective interaction between two strongly adsorbed stars is based on the Daoud-Cotton
blob picture \cite{daoud82} of multiarm stars and its generalization to two interacting stars \cite{pincus.mm.1986},
appropriately adopted to two spatial dimensions. Each of the $f$ arms around the center of
a star can be considered as being built of a succession of concentric blobs of radius $\xi(s)$,
where $s$ denotes the distance from the star center. The polymer segment confined within
each blob is performing a two-dimensional, self-avoiding random walk (SARW), whereas the blobs are
space-filling within the star corona. The latter assumption leads to the scaling $\xi(s) \cong s/f$;
the former implies that the number of monomers, $g(s)$, from a single chain at distance
$s$ from the center, scale as $g(s) \sim \xi^{1/\nu}(s)$, where $\nu = 3/4$ is the SARW-critical exponent
in two dimensions. Concomitantly, the monomer concentration $c(s)$ scales with the distance
from the star center as:
\begin{equation}
c(s)\simeq \frac{f \xi^{1/\nu}(s)}{2\pi s \xi(s)}\sim f^{2-1/\nu}s^{1/\nu-2}a^{-1/\nu},
\label{rhor}
\end{equation}
where $a$ denotes the segment size.
Using the value of the scaling exponent $\nu$=3/4 
in two dimensions under good solvent conditions \cite{benhamou03},
one obtains:
\begin{equation}
c(s)\simeq f^{2/3} \left(\frac{a}{s}\right)^{2/3} a^{-2}.
\label{rhoscale}
\end{equation}
The size of the star $R$, standing for, e.g., its gyration radius $R_g$ or its center-to-end radius $R_e$,
can be determined by the conservation law of the monomer number,
 $fN=2\pi\int_{0}^{R} c(s) s {\rm d}s$. The scaling of $R$ follows: 
\begin{equation}
R\sim a f^{1/4}N^{3/4},
\label{rgstar}
\end{equation}
featuring a stronger stretching effect of the arms with $f$ in comparison to the
three-dimensional case \cite{daoud82}, in which $R \sim a f^{1/5}N^{3/5}$. 
The above relations for $c(s)$ and $R$ imply that the 
scaled monomer concentration
$R^2c(s)/(Nf)$ is a universal function $\rho(x)$ of $x \equiv r/R$,
with the properties:
\begin{equation}
2\pi\int_0^1 \rho(x) x {\rm d}x = 1\,\,\,\ {\rm and}\,\,\, \rho(x) \sim x^{-2/3}\,\,{\rm for}\,\,x \leq 1.
\label{rhobar}
\end{equation}
%\begin{figure}[ht]
%\includegraphics[scale=0.3]{tauVSN-eps-converted-to.pdf}
%\hspace{0.7cm}
%\includegraphics[scale=0.34]{tau_N_MC-eps-converted-to.pdf}
%\caption{}
%\label{fig_tauVSN}
%\end{figure}

Our MD simulation results in \ref{fig2}
fully confirm this picture, since density profiles from three different stars
collapse onto one another when properly scaled, showing in addition the power-law
predicted by Eq.~(\ref{rhobar}). 
Furthermore, the validity of Eq.~(\ref{rgstar}) 
is demonstrated in
\ref{fig3}, where the quantity $f^{-1/4}R_g$ is shown to indeed scale as $N^{3/4}$. 
Scaling- and blob-based considerations are thus fully capable of
predicting quantitatively
the properties of strongly adsorbed, individual star polymers.

\begin{figure}
\includegraphics[scale=0.4]{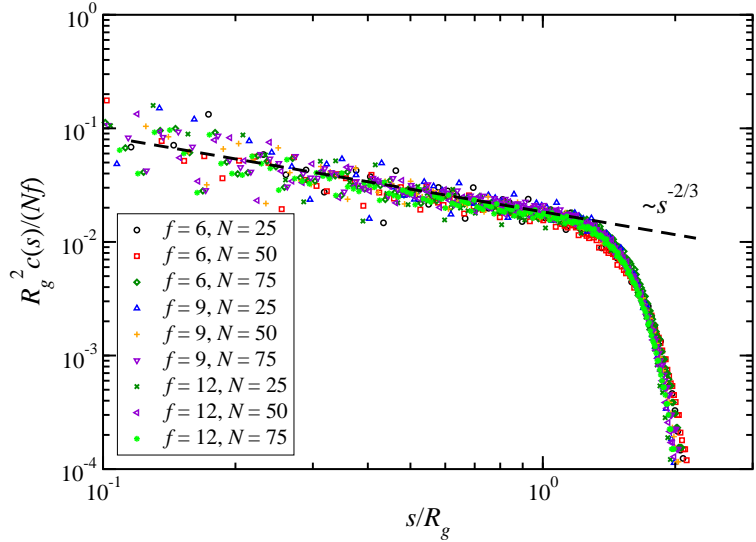}%[width = 8.3 cm]
\caption{MD simulation results for the normalized monomer density
  profiles for various star polymers as a function of dimensionless
  distance from the star center; the dashed line indicates the
  predicted power-law, $c(s) \sim s^{-2/3}$.} 
\label{fig2}
\end{figure}

\begin{figure}
\includegraphics[scale=0.4]{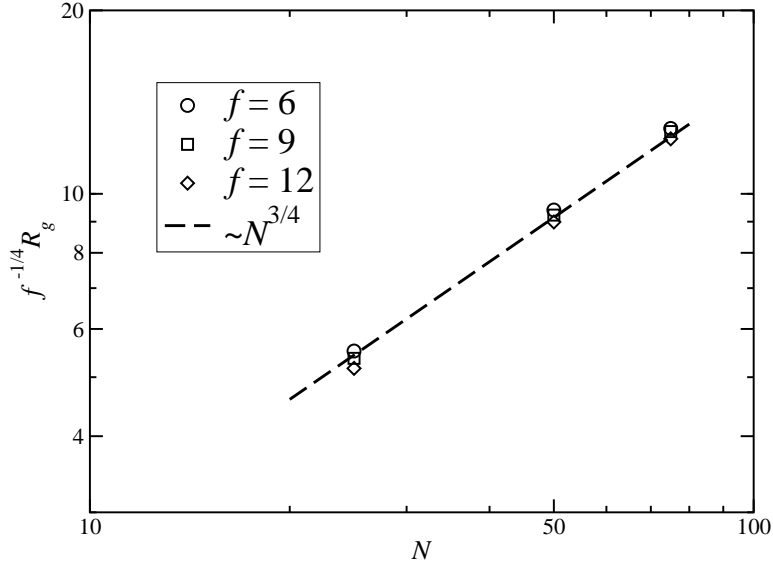}%[width = 8.3 cm]
\caption{Simulation results for the star radius
$R_g$ scaled by $f^{1/4}$ as a
function of the star arm length $N$ for several values of $f$ and $N$;
 the dashed line indicates the power law of Eq.~(\ref{rgstar}).} 
\label{fig3}
\end{figure}

The theoretical approach to the functional form of the effective interaction
$\beta V_{\rm eff}(d)$ between two star polymers in two dimensions, whose anchoring
points are separated by distance $d$, proceeds along two parallel lines,
one pertaining to the case $f=1$ (two polymer chains) and the second to the
case $f \gg 1$ (multiarm stars). For $f=1$, we consider first the case of a single,
self-avoiding chain of $N$ segments, which possesses 
a partition function $Z_1(N)$ that scales as $Z_1(N) \sim \tau^{N}N^{\gamma-1}$ \cite{pincus.mm.1986}.
Here, $\tau$ is a model-dependent parameter and $\gamma = 43/32$ is another 
universal exponent of the 2d-SARW \cite{benhamou03}. Two such stars, when kept at infinite separation,
are thus characterized by a partition function $Z_2(d \to \infty) = Z_1^2(N) \sim \tau^{2N}N^{2(\gamma-1)}$.
When their end-monomers are kept at separation $r \sim a$, the two chains resemble
a single SARW of $2N$ steps, thus $Z_2(d \sim a) \sim \tau^{2N}(2N)^{\gamma-1}$.
On the other hand, the segment length can be renormalized to larger values as long
as it does not exceed, roughly, the equilibrium chain size $R$.
The effective interaction is given as $\beta V_{\rm eff}(d) = -\ln[Z_2(d)/Z_2(d \to \infty)]$.
Using the above considerations in conjunction with the relation $R \sim N^{\nu}$,
we obtain
\begin{equation}
\beta V_{\rm eff}(d) = -\frac{\gamma - 1}{\nu}\ln\left(\frac{d}{R}\right)\,\,{\rm for}\,\,d \lesssim R,
\label{veff_chain:eq}
\end{equation}
for the case $f=1$, where $(\gamma -1)/\nu = 11/24$.

In the opposite case, $f \gg 1$, we proceed by first counting the number of Daoud-Cotton
blobs, $M(f)$, of a single star. The condition $c(s) \sim \xi^{1/\nu}(s)$
together with the geometrical picture of closely-packed blobs lead, after some
straightforward algebra, to the result $M(f) \sim f^2\ln(R/R_c)$, where $R_c$ is a
(microscopic) scale denoting the size of the core of the star, on which the chains
are grafted; this result should be compared with the corresponding scaling
$M(f) \sim f^{3/2}\ln(R/R_c)$ in three dimensions \cite{pincus.mm.1986}, demonstrating, once more,
the stronger stretching of the chains in two dimensions. If two such multiarm stars
are now brought to a separation of order $R_c$ from one another, they resemble 
a single star with $2f$-arms, forming a number $M(2f) \sim (2f)^2\ln(R/R_c)$ blobs.
The effective interaction can be estimated by assigning the usual free energy cost
of $k_{\rm B}T$ per blob, i.e., $\beta V_{\rm eff}(d) \sim M(2f) - 2M(f)$, leading to
the expression:
\begin{equation}
\beta V_{\rm eff}(d) = -\Theta(f)\ln\left(\frac{d}{R}\right)\,\,{\rm for}\,\,d \lesssim R,
\label{veff_star:eq}
\end{equation}
where the amplitude scales as $\Theta(f) \sim f^2$ for $f\gg 1$.
 
Since a single chain ($f=1$) can be seen as a special case of a star polymer
and the short-range behavior of the effective potential is logarithmic with $r$
in both cases, it is reasonable to request that $\Theta(f = 1) = 11/24$ 
and introduce a general expression for this amplitude that smoothly interpolates
between the two extreme cases. Field-theoretical arguments
lead to the expression \cite{benhamou03}:
\begin{equation}
\Theta(f) = \frac{2 + 9f^2}{24},
\label{theta:eq}
\end{equation}
which, evidently, satisfies the requirements on both limits, $f \to 1$ and $f \gg 1$.
The ultrasoft logarithmic repulsion
for the star-star interaction potential has been obtained earlier {\it in three dimensions} 
on a similar basis, with the
corresponding short-range repulsive force behaving as 
$(5f^{3/2}/{18})\ln(R/d)$ \cite{likos98}. The 
functional form of the short range star-star interaction remains
the same but its amplitude increases
  due to confinement from $5f^{3/2}/{18}$ to 
 $(2+9f^2)/24$ \cite{benhamou03}. 

The effective force $F_{\rm short}(d)$ for $d \lesssim R$ follows as minus  the gradient
of $V_{\rm eff}(d)$. Taking into account that for $d \to R_c$ the strong steric interaction between
the cores dominates over the universal, logarithmic form of the latter arising from
polymer self-avoidance, we are led to introduce a shift $d \to d - R_c$ on the coordinate,
as previously done also for three-dimensional star polymers \cite{sebastian.mm.2009}, yielding:
\begin{equation} 
\frac{F_{\rm short}(d)}{k_BT}=\frac{2 + 9f^2}{24(d-R_c)}.
\label{fshort}
\end{equation} 
In the long-range regime, Flory theory yields a Gaussian form for the
potential of mean force, which results into the following scaling expression
for the mean force at long ($d > R_{e}$) star-star
separations \cite{cerda03}:
\begin{equation} 
\frac{F_{\rm long}(d)}{k_BT}=\frac{2+9f^2}{24(R_{e}-R_{c})}\frac{d}{R_{e}}
\exp\left[-\left({d}/R_{e}\right)^2\right].
\label{flong}
\end{equation} 
Here, the prefactor has been determined using the 
physical requirement of continuity between the short-
and long-range forces at $d=R_{e}$. The corresponding results
for the mean force between two adsorbed stars are shown in
\ref{fig4} together with the simulation data for several values
of $f$ and $N$; the agreement between theory and simulation is
excellent for all values of parameters studied, without use of
fit parameters. 

Setting $x \equiv d/R_e$ and taking the limit $R_c \to 0$ (long chains), we
put forward
a simple and accurate analytical form for the effective potential between two
star polymers in two dimensions as:
\begin{equation}
\beta V_{\rm eff}(x)=\frac{2 + 9f^2}{24}\begin{cases}
-\ln x + {\rm e}^{-1}/2, & \text{if }x \leq 1,\\
{\rm e}^{-x^2}/2, & \text{if } x > 1. 
\end{cases}
\label{veff2d:eq}
\end{equation}
Eq.\ (\ref{veff2d:eq}) extends the hitherto known three-dimensional result \cite{likos98}
to two dimensions, introducing an easy to handle functional form for the latter.
The ultrasoft potential of a logarithmic form is 
a universal feature also for the interaction of dissimilar star polymers \cite{mayer.mm.2007} or
between star polymers and flat or curved walls \cite{daniela.jcp.2012} in three dimensions.
The criterion of a `weak mean field fluid' \cite{likos.pr.2001}, expressed through the condition
$\hat V \equiv \int_0^{\infty} x^{d-1} V_{\rm eff}(x) {\rm d}x < \infty$, is fulfilled for the
effective interaction of Eq.\ (\ref{veff2d:eq}). 

\begin{figure}
\includegraphics[scale=0.5]{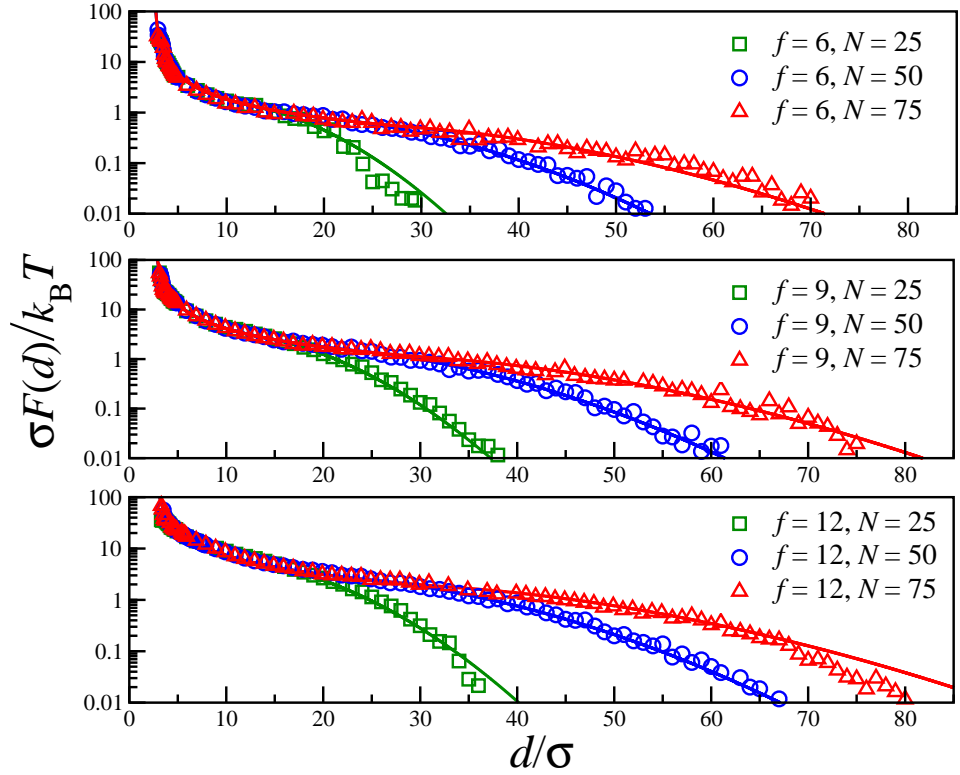}
\caption{MD simulation (symbols) and scaling theory (lines) results
  for the mean force between two adsorbed stars for several values of
  $f$ and $N$, as indicated in the legends.} 
\label{fig4}
\end{figure}

\begin{figure}
\includegraphics[scale=0.5]{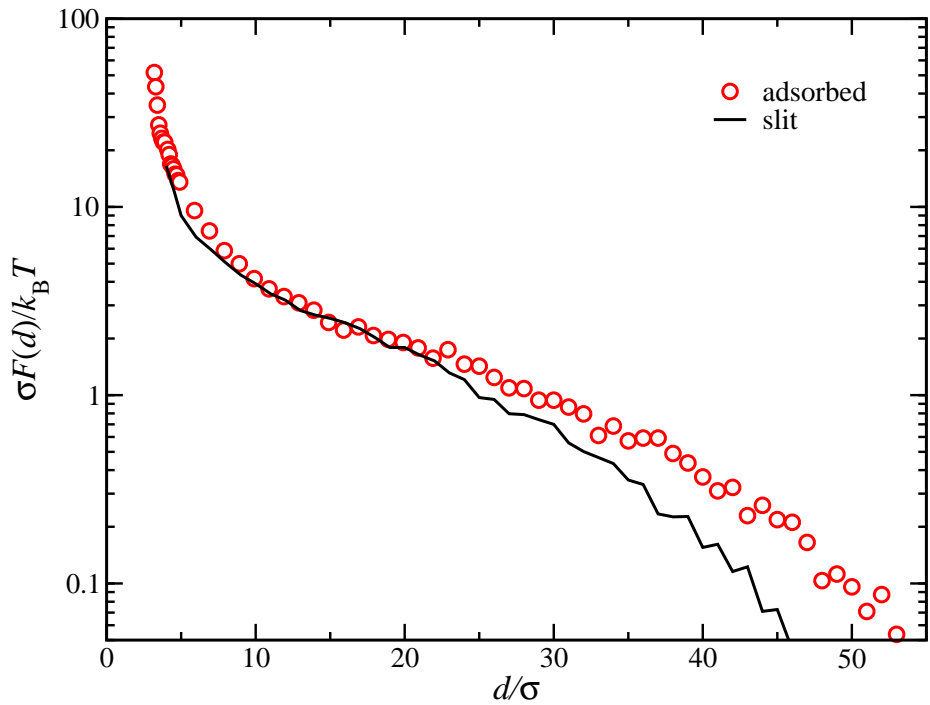}
\caption{Comparsion of MD simulation results for the mean force between two stars adsorbed onto surface (circles) or
trapped in a slit (solid line).
Here  $f=9$ and $N=50$. Slit width is $4\sigma$.} 
\label{fig5}
\end{figure}

Finally, in \ref{fig5} we compare our simulation results for the two distinct cases already
mentioned above:
stars strongly adsorbed on a single wall and stars trapped between two parallel, repulsive walls
forming a slit of width $H = 4\sigma$ in the $z$-direction.
The data indicate a systematic deviation of the two forces, 
with growing inter-star separation distance $d$.  As long as the two stars are
far apart in the slit, the arms on their periphery experience
very low monomer density in their vicinity, and can easily slip above or
beneath one another. This is reflected by the smaller force of
repulsion, felt by the stars, compared to the case of strong adsorption.
In the latter case, the arms cannot intersect as they stick strongly to
the substrate, so they either have to bend when encountering an arm of
the second star (meaning stronger repulsion), or crawl over it which is
less probable because it costs energy of contacts with the substrate. On
the other hand, the forces coincide at smaller distances $d$ because the
increased density around the star centers largely prevents easy
threading of arms through each other, so that there is no way to avoid
the strong repulsion.

\section{Conclusion}
We have demonstrated by a combination of molecular simulation and analytical theory 
that soft colloids, represented here by the common example of star polymers, experience dramatic
conformational changes upon confinement in two dimensions, which profoundly affect their effective
interaction, rendering it significantly more repulsive than in three dimensions.
Previous studies on three-dimensional star polymers \cite{ferber} indicated that the pair forces continue 
to provide a valid and accurate description of the interactions even for the case in which three 
stars approach each other at distances that represent a small fraction of their respective sizes.
Thus, our results for effectively two-dimensional star polymers are expected to be quantitatively reliable 
for concentrations well above the overlap density of the stars.
In addition our checks have shown that effective interaction between two adosorbed stars is the same
as the effective force for stars trapped in a very narrow slit formed by two paralell and repulsive walls.
Accordingly, we expect 
that much stronger ordering will take place in confinement and that the structural arrest lines will
shift to lower concentrations with respect to their bulk values. Changing the amplitude of the attractive
interaction to the wall allows us to tune the conformations and the interactions of the (partially) adsorbed
stars, opening a pathway to tunable patterning of surfaces with ordered arrays of adsorbed stars.

Possible experimental validation of our threoretical predictions would be to graft long chains on colloidal particles
and use optical tweezers in order to measure effective force. The alternative approach would be to measure the correlations
between the adsorbed stars in the same way this has been done for other 2d-colloidal systems adsorbed on surfaces
or interfaces (see e.g. \cite{Hoffmann}). 
The resulting correlations should then be describable by the effective interactions derived in this paper.

Our work opens up a possibility to consider also the case of
partially absorbed stars with $f_a$ chains lying on the plain and $f-f_a$ chains assuming
3d-configurations, for which the force can be obtained as the sum of the 2d- and 3d-expressions,
each weighted with the appropriate amplitude.
The determination of the interaction between partially adsorbed stars, bridging the gap
between two and three dimensions, will be the subject for future work.

\section*{Acknowledgements}
S.A.E. acknowledges financial support from the Alexander von Humboldt
foundation. A.M. gratefully acknowledges the hospitality and financial support by CECAM nano SMSM at the Max-Planck Institute for Polymer Research in Mainz, 
Germany, during this study. This work was partially supported by the Marie Curie ITN-COMPLOIDS (Grant Agreement No. 234810).
Computational time on the PL-Grid Infrastructure is gratefully acknowledged.

%\newpage
%\begin{center}
%{\bf \Large Controlling the Interactions between Soft Colloids via Surface Adsorption }\\
%\vskip 0.2 true cm
%\end{center}
%
%by Sergei A. Egorov, Jaros{\l}aw Paturej, Christos N. Likos and  Andrey Milchev
%
%
%
%TOC graph:
%\begin{figure}[ht]
%\includegraphics[scale=0.49]{toc.eps}
%\end{figure}
%\begin{center}{\it Simulation snapshot of two strongly interacting 
%adsorbed star polymers.}\end{center}
%
\end{document}